# Phishing and Spear Phishing attacks: examples in Cyber Espionage and techniques to protect against them

Alessandro Ecclesie Agazzi
*Department of Computing and Informatics*
Bournemouth University
Poole, United Kingdom
alessandro@ecclesieagazzi.com

*Abstract*— Phishing attacks have become the most used technique in the online scams, initiating more than 91% of cyberattacks, from 2012 onwards. This study reviews how Phishing and Spear Phishing attacks are carried out by the phishers, through 5 steps which magnify the outcome, increasing the chance of success.

The focus will be also given on four different layers of protection against these social engineering attacks, showing their strengths and weaknesses; the first and second layers consist of automated tools and decision-aid tools. The third one is users' knowledge and expertise to deal with potential threats. The last layer, defined as "external", will underline the importance of having a Multi-factors authentication, an effective way to provide enhanced security, creating a further layer of protection against Phishing and Spear Phishing.

## I. INTRODUCTION

Originally invented in 1989 to connect universities and scientific communities around the globe [1], the Internet is now a popular tool for entertainment, to run activities, communicate with friends and deliver information. However, some people are exploiting their anonymity to cheat others; Phishing attacks are the most widely used technique [2].

Phishing has been defined as a "*social engineering attack that uses e-mail, social network webpages, and other media to communicate messages intended to persuade potential victims to perform certain actions* [e.g. entering login credentials in a cloned webpage, downloading an attachment embedded with a malware or opening an infective hyperlink] *or divulge confidential information for the attacker's benefit in the context of cyber security*" [3].

Phishing can lead to the exposure of intimate information, financial loss and a reduction in trust [4]. More than 91% of cyberattacks, from 2012 onwards, were initiated with a Phishing attack and in 2018 almost one third of the total data breaches involved Phishing or Spear Phishing [5], which is a targeted Phishing attack, studied and applied to a specific user or set of users [6].

This paper aims to investigate how Phishing and Spear Phishing attacks are performed, by analysing 5 different steps a Phisher carries out to hack the victim. Furthermore, different techniques which help users to protect against them will be presented: automated tools and decision-aid tools are the first two layers of defence.

The remainder of this paper will point out how knowledge and training, which constitute the third layer of protection, are the most effective techniques not to be deceived by Phishing.

In conclusion, the Multi-Factor Authentication (MFA), a security system which certifies users' identity by requesting multiple credentials will be suggested as possible further and "external" layer of defence.

## II. SPEAR PHISHING AND SOCIAL MEDIA

Although e-mail remains the favourite medium for the Spear Phishing attacks, social media has now opened the way to new attack vectors [7]. These "social" platforms are the most effective place for scammers to devise elaborate Spear Phishing campaigns, often against politicians and the military [8]; indeed, 25% of all Phishing attacks are against state actors and 70% of all state attacks are by Phishing [9].

The 2016 U.S. election can be considered a 'classic' example of cyber espionage by a Spear Phishing attack. John Podesta, Hillary Clinton's campaign manager, and Colin Powell, the former U.S. Secretary of State, were deceived by fake phishing advertisements and their credentials were stolen by the Russian hackers' group, Fancy Bear. All the mails between them and Hilary Clinton were then published on the webpage WikiLeaks [10].

A problem, such as this, could be not only experienced by state actors but also by the general public, who potentially are less protected.

## III. A SPEAR PHISHING MODEL

The 'theoretical model of Spear Phishing on social media' developed by Michael Bossetta shows how illiberal regimes perform this kind of social engineering attack to assault Western democracies [11].

Although Michael Bossetta specific context was "political", it could be argued that his theory is more widely applicable to the general public. This theory involves 5 steps through which the Phisher leads the spear-phishing attack:

## 1. Collect

Gathering information on a specific target by exploiting all the available information shared throughout social media to increase the chances of a successful attack [12]. Actors have, at this point, collected personal information such as phone numbers, e-mail addresses, interests, working or educational history [13].

In the 2017 French election, Facebook detected and blocked 24 profiles collecting information about Macron's campaign. The actor behind those profiles was observing the victims in order to draw up an elaborated and customised attack. The "bad actor", using Bossetta's term, was discovered to be the same one as in the American election, the Russian hacking group Fancy Bear [14].

## 2. Construct

Creating fake social media accounts in order to engage with targets. The new persona usually has fabricated credentials which establish a common ground with the victim, very often, the same work organisation or university [15].

A Spear Phishing attack by Iranian hackers hit the Saudi Arabian organizations, by creating fake LinkedIn, Facebook, Twitter and Instagram accounts; "Mia Ash" which mimicked an appealing Romanian female, sharing common interests with the victims [16].

## 3. Contact

Contacting the victims. The first approach begins with the "friend request" and once accepted, the phisher can have access to non-public information about the target [17]. Another way of contact is the direct message which could contain the infected hyperlink. In the American election, mentioned above, the method of contact used by the Russian group was the advertising campaign. Using the algorithm, they calculated a strategic time when people were mostly online, to send tweets likely to be clicked by their victims. The attack was successful, 70% of people clicked on those advertisements [18].

## 4. Compromise

Installing malware on the victim's device in order to compromise it and steal information [19]. This is what CyberCaliphate, the Isis-affiliated group did in 2015, they hacked many Youtube and Twitter accounts of the U.S. Central command's employees to spread propaganda [20].

Once malware is installed, the victim may be unaware, and attackers might remain undetected for years [21].

## 5. Contagion

The last step of this model is contagion. Bad actors could exploit the hacked account to spread the attack over related people, magnifying the outcome [22].

As Bossetta explains: "*Contagion is especially dangerous because threat actors can target vulnerable victims and scale up to bigger targets*" [23].

In the case of the U.S. State Department, an Iranian hackers' group in 2015 managed to violate the administration department by compromising the Facebook accounts of young employees [24].

## IV. HOW TO DEFEND AGAINST THE ATTACKS

The protection against the Phishing and Spear Phishing attacks is provided through different layers of defence:

### 1. Automated tools

Prevention and detection are the first layer of protection against Phishing attacks, thus different automated tools for Phishing detection have been implemented [25]:

- Anti-phishing e-mail tool on a server to block Phishing content.
- Websites containing blacklist of detected Phishing URLs and IP addresses.
- Browser extensions to assess whether a page is genuine or not (Phishing websites).

Anti-phishing e-mail tools are meant to block Phishing content such as link, contained in mails, at server and client levels and it is estimated that the 89.5% of infected mails are detected through this process. [37]

Furthermore, websites containing blacklists of detected Phishing URLs and IP addresses are used to assess whether a webpage is genuine or not in case the user encounters a possible scam link which asks for credentials or bank information.

Many Browsers provide extensions meant to alert people about a not genuine page. They are based on blacklist databases of old detected fraudulent URLs, regularly updated.[38]

However, although researchers have suggested these techniques to prevent Phishing attacks, Phishers are becoming ever-more sophisticated in their approaches following the structure, explained above, and creating strategies in order to bypass the Anti-Phishing tools [26].

### 2. Decision-aid tools

In the context of e-mails, e-mails filter tools are not infallible, whenever the infected e-mails step over the detection tool, people are called to make a decision concerning a webpage or link's legitimacy [27].

As a result, decision-aid tools were introduced as a second layer of protection in order to alert people regarding a detected fraudulent website:

- Dynamic security skins.

    A tool which allows an external server to demonstrate the integrity of the questionable link without compromising the machine and the information of the user [42].

- Browser toolbars.

    A Browser toolbar meant to provide feedback (red icon when the link is infective, orange if it is unknown or green if secure) upon the legitimacy of the web page the user has opened [41].

- Browser Phishing notifications.

    These work like the above tools (Browser toolbars) but the feedback given to the user is a "flash notification", which makes him think about the integrity of the page and its provenance.

- Sockets layer warnings.

Although many Phishing attacks are carried out over HTTP protocol, a significant number run on websites for which SSL certificates have been issued [39]. Sockets layer warnings alert users regarding an infective link running on the protocol HTTPS which should be, "theoretically", secure [40].

These tools identify potential risks, actively or passively, to online user [28]:

- Active warnings: coloured graphic icons that catch user's attention, by making users take action regarding the danger (either closing the webpage or continuing).

- Passive warnings: coloured graphic icons which notify users about a possible danger without blocking their activity.

3. *Training and knowledge*

"*Yet, the decision-aid tools have been evidenced as ineffective with usability issues. Users showed lack of understanding of the decision aid warnings in general*" [29].

"*The last line of defence is your human firewall*" [30], hence training and knowledge have proved to be the most useful and crucial layer of protection; prior researches, indeed, showed that knowledge gained from training enhanced effectiveness of a Phishing warning making people act safely[31].

4. *External protection*

A part these 3 layers, another form of protection could be provided by Multi Factor Authentication (MFA), a security system which certifies users' identity by involving multiple credentials [44].

The Single factor authentication, which implies the user to enter only his credentials (usually mail or username and password) in order to get the access to either a website or system, has to be replaced by two-factor authentication (2FA) or better multi-factor authentication (MFA) [43].

Indeed, rather than just asking for a username and password, MFA needs other credentials, e.g. fingerprint, facial recognition or an answer to a pre-set question [48].

In cases where the bad actor is attempting to obtain user login credentials, multi-factor authentication (MFA) has proved to be a key defence. In the worst-case scenario MFA mitigates the malicious actor's ability to log in to a given account or server, limiting the damage [34].

MFA is an effective way to provide enhanced security, creating a further layer of protection. Traditional usernames and passwords may be stolen by the Phisher and scammer, through Phishing and Spear Phishing [47]. Moreover, credentials could be found through the brute force attacks or be already published on the deep web, visible to anyone looking for databases breaches.

Here below there are some examples of different MFA:

- Codes generated by mobile applications.
- Codes generated by mobile USB.
- Fingerprints.
- Codes sent via SMS or email.
- Facial recognition.
- Iris scan
- Answer to pre-set personal questions

What is more, MFA functionality refers to 3 different types of "authentication factors" [45]:

- What you are: authenticated by biometrics, such as fingerprints or iris scan.
- What you have: authenticated by a code sent via mail or SMS.
- What you know: authenticated by a password or PIN.

Multi-factor authentication does not offer a foolproof protection, but the combination of more than one authentication makes hacker's work harder to get into users' account.

V. CONCLUSION

After the disclosure of the Cambridge Analytica scandal and the Russian interference in the America election of 2016, social media platforms removed more than one thousand accounts related to the manipulation in order to limit the hacking activity [32].

In addition, several Anti-Phishing tools have been implemented and Decision-aid tools were introduced in order to help people make the right choice when they come across a possible Phishing danger.

Different examples of Spear Phishing used for political reason, given in this paper, demonstrate that every network, even the most protected and sensible ones, can be exploited and hacked. Social engineering with Phishing and specifically Spear Phishing attacks are based on Human emotionality, making their detection more difficult.

Physical defence layers are, undoubtedly, useful but not enough [46]; people have to be trained in order to achieve the awareness that not every page or link is genuine, everything has to be questioned and assessed, even a message from a person who pretends to be a friend [33].

In addition, as the recent Hacking Humans podcast suggests, attackers hold advantage over the users: "*It's not like a military operation, where the defender is meant to have most of the advantages. In cyberspace, the attacker can just keep trying and probing at low risk and low cost and he only has to be successful once. And e-mails filters designed to keep malicious scams out have a 10.5% failure rate. That sounds pretty good but is not 0%. This isn't baseball, if your technical defence fails in 1 out of 10 tries, you are out of luck and business. The last line of defense is your human firewall*" [35].

To conclude, a constant and updated training helps people to be less vulnerable to these attacks, changing the fortunes of this "cyber war". MFA may also enhance users defence layer in case their credentials have already been stolen, becoming the "very" last wall of protection.